\documentclass[12pt]{article}
\usepackage{amsmath,graphicx}

\date{\today}

\begin{document}
\author{Roberto Brambilla$^1$,Francesco Grilli$^2$,Luciano Martini$^1$ \\
$^1${\small RSE -- Ricerca sul Sistema Energetico, Italy} \\
$^2${\small KIT -- Karlsruhe Institute of Technology, Germany}
}
\title{Critical state solution and AC loss computation of polygonally arranged thin superconducting tapes}

\maketitle

\begin{abstract}
The current density and field distributions in polygonally arranged thin superconducting tapes carrying AC  current are derived under the assumption of the critical state model. Starting from the generic Biot-Savart law for a general polygonal geometry, we derive a suitable integral equation for the calculation of the current density and magnetic field in each tape. The model works for any transport current below $I_c$, which makes it attractive for studying cases of practical interest, particularly the dependence of AC losses on parameters such as the number of tapes, their distance from the center, and their separation.
\end{abstract}
High-temperature superconducting (HTS) cables have a rather complex geometry, generally consisting of several layers of tapes wound around a cylindrical former. Usually the winding angle is relatively small (typically less than 20 degrees), which means that, as a first approximation, the axial component of the produced field can be neglected and the cable can be simulated only with its two-dimensional cross-section, which consists of tapes polygonally arranged around the former. In spite of their simplicity, simple cables made of just one layer of HTS tapes are structures very interesting to investigate, because they can represent several situations: for example, one phase in a three-phase cable, short power links used to connect superconducting devices, or prototypes of more complex cables. 

Calculating the current or field distributions and the AC losses for such geometry can be done with any of the numerical models for superconductors that have been recently developed, following various approaches and formulations~\cite{Grilli:TAS13b}. However, they usually require lengthy software implementations. For this reason, here we derive an analytical solution of the problem, based on the critical state. This work follows and completes earlier works: using conformal transforms, Norris~\cite{Norris:JPDAP70} derived an expression (later corrected by Majoros~\cite{Majoros:PhysC96}) for the losses of an infinitely thin plane with a slit; Mawatari~\cite{Mawatari:APL06}, also using conformal transforms, developed an analytical model for the very same geometry considered here, but based on the Meissner state: as a consequence, his results are valid only for very small transport currents -- a condition seldom met in practical applications. In contrast, in the present model we take into account the penetration of the transverse magnetic field at the rims of the tapes
(where the current density assumes the saturation value $J_c$) and derive an appropriate integral equation
whose solution, for any penetration depth, gives the corresponding current density, and, by the 
Biot-Savart law, the penetrated magnetic field. From the latter the AC losses can be evaluated without
the exceedingly restrictive assumption of a low value of the transport current.

\begin{figure}[t!]
\centering
\includegraphics[width=8 cm ]{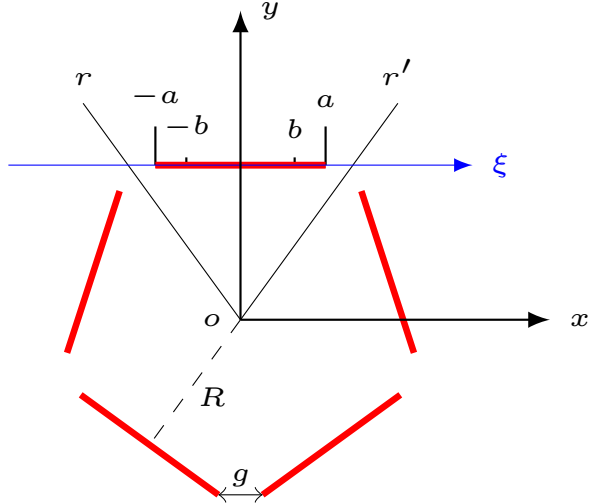}
\caption{\label{fig:penta_plot}Schematic representation of five polygonally arranged superconducting tapes.}
\end{figure}

Let us consider a system of $n$ superconducting tapes of width $2a$ symmetrically lying in a polygonal arrangement, as shown in Fig.~\ref{fig:penta_plot}. In order to avoid tape overlapping, the condition $R>a\cot(\pi/n)$ must hold. Because of the symmetry we can just consider the sector $ror'$, whose representing tape is the segment $(-a,a)$ on thehorizontal axis $\xi$ placed at a distance $R$ from the center. 
Using the Biot-Savart formula for a system of currents with angular periodicity $2\pi/n$, the magnetic field is~\cite{Brambilla:TAS12}
\begin{equation}
H(z)=H_y(z)+iH_x(z)=\frac{1}{2\pi}\int\limits_{-a}^a J(\xi)\frac{nz^{n-1}}{z^n-(\xi+iR)^n}d\xi,
\end{equation}
where the apothem $R$, for a gap g between adjacent tapes, is given by  $R=a/\tan (\pi/n) + g/(2\sin (\pi/n))$.
Based on the critical state theory, at the first peak of magnetic field the current density reaches the critical value $J_c$ in two bands $(-a,-b)$ and $(b,a)$, whereas in the central region $(-b,b)$ it assumes a distribution $g(\xi)$ to be found

\begin{align}\label{eq:slab}
J(\xi)=J_c \times
\begin{cases}
1  & (-a<\xi<-b) \\
g(\xi)  & (-b<\xi<b) \\
1  & (b<\xi<a).
\end{cases}
\end{align}
The transversal magnetic field $H_y=\Re (H)$ is null in the central zone, so, considering only the real part, we have
\begin{equation}
0=H_y(x+iR)=\int\limits_{-a}^a J(\xi) \Re \left [\frac{1}{2\pi}\frac{n(x+iR)^{n-1}}{(x+iR)^n-(\xi+iR)^n}\right]d\xi,
\end{equation}
or, breaking the integral over the three intervals $(-a,-b)$,$(-b,b)$,$(b,a)$,
\begin{equation}\label{eq:Cauchy}
{\rm p.v.}\int\limits_{-b}^b g(\xi) \Re \left [\frac{1}{2\pi}\frac{n(x+iR)^{n-1}}{(x+iR)^n-(\xi+iR)^n}\right]d\xi = f(x)
\end{equation}
where the known term is 
\begin{align}
\nonumber
f(x)&=-J_c\int\limits_{-a}^{-b} + \int\limits_{b}^a \Re \left [\frac{1}{2\pi}\frac{n(x+iR)^{n-1}}{(x+iR)^n-(\xi+iR)^n} d\xi \right] \\
\nonumber
&=\Phi_n\left( \frac{iR-a}{iR+x}\right) 
-\Phi_n\left( \frac{iR+a}{iR+x}\right) \\
\label{eq:known_term}
&+\Phi_n\left( \frac{iR+b}{iR+x}\right)
-\Phi_n\left( \frac{iR-b}{iR+x}\right)
\end{align}
where
\begin{equation}
\Phi_n(z)=J_c\frac{n}{2\pi}z~_2F_1(1,1/n;1+1/n;z^n).
\end{equation} 
Equation~\eqref{eq:Cauchy} is a singular integral equation of Cauchy type, because the integral has to be taken as principal value. In order to solve it, it is necessary to extract its singularity; one can do that by using the identity $x^n-y^n=(x-y)\lambda_n(x,y)$, where $\lambda_n(x,y)=\sum_{k=0}^{n-1}x^{n-1-k}y^k$. Then one has
\begin{equation}
{\rm p.v.}\int\limits_{-b}^b \frac{g(\xi)}{x-\xi} \Re \left [\frac{1}{2\pi}\frac{n(x+iR)^{n-1}}{\lambda_n(x+iR,\xi+iR)}\right]d\xi = f(x)
\end{equation}
and, defining the dimensionless quantity
\begin{equation}
Q(x,\xi)= \Re \left [ \frac{1}{2\pi}\frac{n(x+iR)^{n-1}}{\lambda_n(x+iR,\xi+iR)}\right],
\end{equation}
one can extract the singularity
\begin{equation}
Q(x,x){\rm p.v.}\int\limits_{-b}^b \frac{g(\xi)}{x-\xi} d\xi+ \int\limits_{-b}^b \frac{Q(x,\xi)-Q(x,x)}{x-\xi}g(\xi)d\xi= f(x).
\end{equation}
From the definition of $\lambda_n(x,y)$, it follows that $Q(x,x)=1/2\pi$, so we can write
\begin{equation}\label{eq:2piFx}
{\rm p.v.}\int\limits_{-b}^b \frac{g(\xi)}{x-\xi} d\xi +\int\limits_{-b}^bP(x,\xi)g(\xi)d\xi= 2\pi f(x),
\end{equation}
where we  defined the {\it non-singular} kernel
\begin{equation}
P(x,\xi)=\frac{2\pi Q(x,\xi)-1}{\xi-x}.
\end{equation}
 As a matter of fact, using the definition of $Q(x,\xi)$, one can  verify that $P(x,x)=(n-1)x/2(x^2+R^2)$.
 
Equation~\eqref{eq:2piFx} can be numerically solved without difficulty, by transforming it into an equivalent non-singular Fredholm equation of the second kind (see Appendix)
\begin{equation}\label{eq:gx}
g(x)+\int\limits_{-b}^bK_0(x,\xi)g(\xi)d\xi=f_0(x),
\end{equation}
where
\begin{align}
& K_0(x,\xi)=\frac{1}{\pi^2}\sqrt{b^2-x^2}{\rm p.v.}\int\limits_{-b}^b \frac{P(t,\xi)}{\sqrt{b^2-t^2}}\frac{dt}{t-x} \\
& f_0(x)=\frac{1}{\pi^2}\sqrt{b^2-x^2}{\rm p.v.}\int\limits_{-b}^b \frac{2\pi f(t)}{\sqrt{b^2-t^2}}\frac{dt}{t-x}.
\end{align}
The singularities have in this way been moved from the equation to given functions.
After solving~\eqref{eq:gx} by means of standard numerical methods for integral equations, we can compute the  current transported by each tape and the transversal magnetic field along the tape's width (the $\xi$ axis), respectively
\begin{align}
& I_t=2J_c(a-b)+J_c\int\limits_{-b}^b g(\xi) d \xi  \\
&H_y(\xi)=\int\limits_{-a}^a J(\xi)\Re \left[\frac{1}{2\pi}\frac{n(\xi+iR)^{n-1}}{(\xi+iR)^n-(\tau+iR)^n}\right]d\tau.
\end{align}
\begin{figure}[t!]
\centering
\includegraphics[width=8 cm]{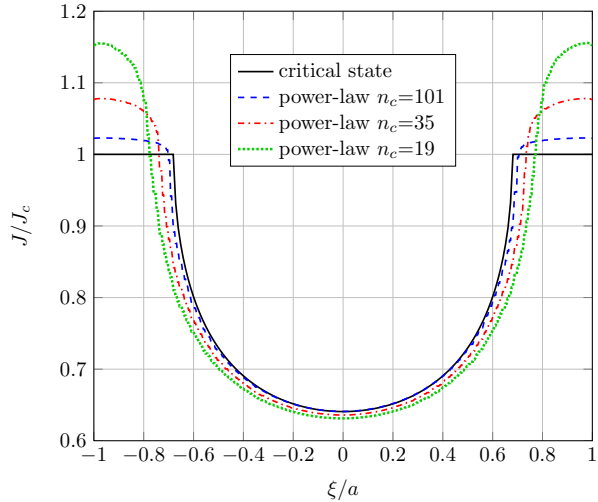}
\caption{\label{fig:fig2}Current density profile along the tape's with for a polygonal arrangement with $n$=5 and gap=0.5 mm. Shown are also the profiles calculated by FEM simulations describing the superconductor with a power-law resistivity with finite exponent.}
\end{figure}

Figure~\ref{fig:fig2} shows  the current density profile along the $\xi$ axis for the  case $n$=5, $a$=6 mm, $I_c$= 360 A, $I/I_c$=0.8 (these values of the tape's width and $I_c$ are typical of state-of-the-art rare-earth-based coated conductors and are used in the remainder of the paper as well). Shown are also the profiles calculated with 2-D finite-element method (FEM) simulations based on the $H$-formulation~\cite{Brambilla:SST07} of Maxwell equations and power-law resistivity $\rho$=$E_c/J_c |J/J_c|^{n_c-1}$ for the superconductor material, where $E_c$=$10^{-4}$ V/m and $n_c$ is the power index associated to the flux creep. The current profiles for a finite $n$-value are also fully matched by the  FEM-based 1-D integral equation model with power-law resistivity for polygonally-arranged tapes~\cite{Brambilla:TAS12}. In the FEM simulations, the frequency of the current source is 50 Hz. 

Due to the utilized power-law, the FEM model allows the current density to exceed $J_c$ and the profiles are smoother than those obtained with the critical state model. When the power index is increased from $n_c$=35 (a typical value for coated conductor tapes) to $n_c$=101, the profiles approach those calculated with the critical state model. 

Once the current and field profiles are known, one can easily compute the AC losses of the $n$-polygonal cable for a given value of $b$, which corresponds to a given value of the transport current, using Norris's method based on the knowledge of the distribution of the transversal magnetic field at the instant of maximum penetration~\cite{Norris:JPDAP70}
\begin{equation}\label{eq:AC_losses}
Q=8\mu_oJ_c \int\limits_b^a (a-\xi)H_y(\xi)d\xi.
\end{equation}

\begin{figure}[t!]
\centering
\includegraphics[width=8 cm]{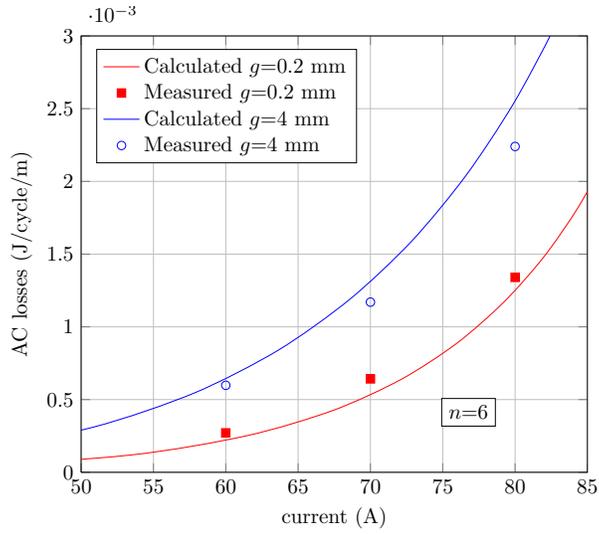}
\caption{\label{fig:Ogawa_plot}Comparison between calculated and measured losses for a polygonal cable: transport AC losses of  hexagonal cables with gaps of 0.2 and 4 mm. Calculations were performed with $I_c$=95 A.}
\end{figure}

\begin{figure}[t!]
\centering
\includegraphics[width=8 cm]{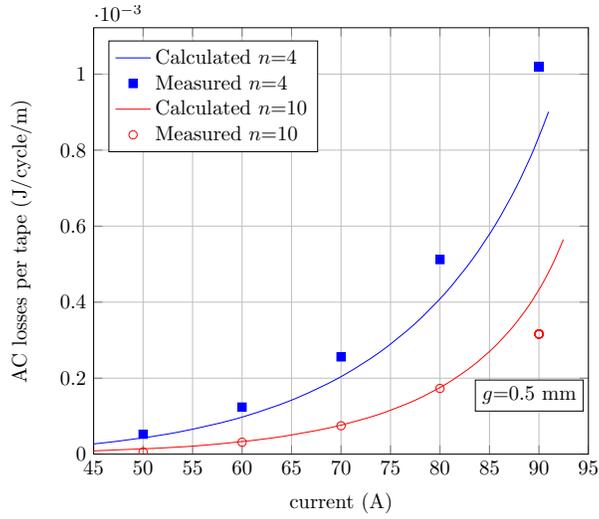}
\caption{\label{fig:Ogawa_plot2}Comparison between calculated and measured losses for a polygonal cable: transport AC loss (per tape) for cables composed of 4 and 10 tapes. Calculations were performed with $I_c$=95 A.}
\end{figure}

In order to further validate our model, we compared our loss results  with the measured transport losses of cables made of polygonally arranged yttrium-barium-copper-oxide (YBCO) tapes connected in series, which ensures that the same current flows in all the tapes~\cite{Ogawa:TAS11}. Figure~\ref{fig:Ogawa_plot} shows the comparison for an hexagonal cable with different values of the gap, whereas Fig.~\ref{fig:Ogawa_plot2} for cables with various numbers of tapes and a fixed gap. The experimental loss data were extracted from Fig. 6 and Fig. 8 of the paper referenced above. In general, the agreement between our calculations and the experimental data is good. Our model correctly predicts the variation of the AC losses with the gap and number of tapes in the cable: reducing the gap reduces the perpendicular magnetic field component impinging on the tapes and hence the AC losses. For a fixed gap value, increasing the number of tapes makes the field lines more concentric, so that the perpendicular field component (responsible for the losses in infinitely thin tapes) is progressively lowered and so are the AC losses. The deviation from experiments can be considered acceptable given the numerous possible causes of discrepancy (in addition to the numerous intrinsic issues related to AC loss measurement~\cite{Tsukamoto:TAS01,Stafiniak:TAS09}): possible $I_c$ variations of the tapes composing the cable, variations of the actual gap in the experiments, effects of $J_c(B)$ dependence (not taken into account by our model).

\begin{figure}[t!]
\centering
\includegraphics[width=8 cm]{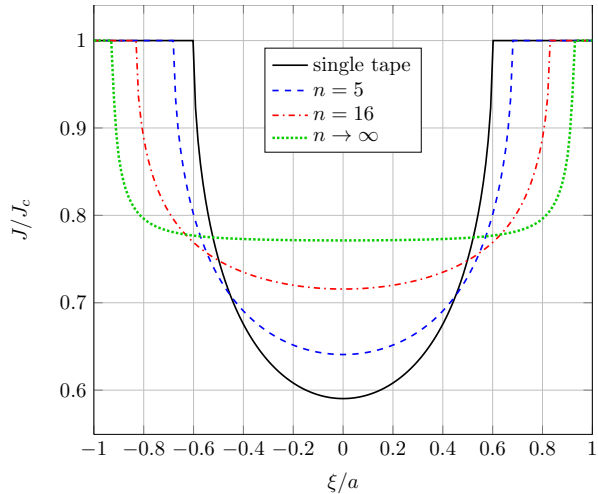}
\caption{\label{fig:J_comparison_plot}Comparison of the current density profiles for a single tape, for a polygonal arrangement with $n$=5, $n$=16, and for the limit $n \rightarrow \infty$ (infinite $X$-array). The gap is 0.5 mm and the applied current $I/I_c$=0.8.}
\end{figure}

{\bf X-array --} When the number tapes becomes very large, the polygonal arrangement approximates that of an infinite $X$-array (infinite array of coplanar strips of width $2a$ separated by a gap $g$); the apothem increases as $R=n L$ where $L=(a+g/2)/\pi$. Inserting this expression in (\ref{eq:Cauchy}) and (\ref{eq:known_term}) and letting $n  \rightarrow \infty$  we easily obtain the kernel and the known term for $X$-array 
\begin{align}\label{eq:K_X}
&K_{\infty}(x,\xi)=-\frac{1}{2\pi L}\Re \left [ \frac{i}{1-e^{i(x-\xi)/L}}\right] \\
\label{eq:f_X}
&f_{\infty}(x)=\frac{1}{2\pi}\Re \log \left [ \frac{e^{i(a-x)/L}-1}{e^{-i(a+x)/L}-1} \cdot \frac{e^{-i(b+x)/L}-1}{e^{i(b-x)/L}-1}\right].
\end{align}
Inserting these expressions in~\eqref{eq:Cauchy} with the variable substitutions $X=\exp(ix/L)$, $A=\exp(ia/L)$, and $B=\exp(ib/L)$, we obtain the integral equation
\begin{equation}
\int\limits_{1/B}^B \frac{J(Y)}{X-Y}dY=\log \left (\frac{A-X}{B-X}\frac{BX-1}{AX-1} \right ).
\end{equation}
\begin{equation}
b=2L \arccos \left [ \frac{\cos(a/2L)}{\cos(pa/2L)}\right ].
\end{equation}

Its solution is~\cite{Polyanin98}
\begin{equation}
J_X(x)=\frac{2}{\pi}\arctan \sqrt{\frac{\tan^2(a/2L)-\tan^2(b/2L)}{\tan^2(b/2L)-\tan^2(x/2L)}},
\end{equation}
which coincides with the result obtained by M\"uller~\cite{Muller:PhysC97a}, with a completely different approach, based on the transformation of the $X$-array to the single tape case. The value of $b$ is determined by imposing the total transport current $I=pI_c$ ($0<p<1$)

Using the kernel~\eqref{eq:K_X} we can obtain the magnetic field inside the tapes
\begin{equation}
H_y(x)=\int\limits_{-a}^a K_{\infty}(x,\xi)J_X(\xi)d\xi.
\end{equation}
and from this, by (\ref{eq:AC_losses}), the AC losses. In Fig.~\ref{fig:J_comparison_plot} we show a comparison of the current density for a single tape, a polygonal arrangement with $n$=5, $n$=16, and $n \rightarrow \infty$ ($X$-array), for $g$=0.5 mm and $p$=0.8. For a fixed gap, increasing the number of tapes reduce the penetration of the current inside the tape, and hence its AC losses. The $X$-array configuration can be useful to simulate one-layer solenoids.

This work was supported by the Research Fund for the Italian Electrical System under the Contract Agreement between RSE and the Ministry of Economic Development (RB, LM)
and by the Helmholtz-University Young Investigator Group Grant VH-NG-617 (FG).

\section*{Appendix}

The singular equation
\begin{equation}\label{eq:singular}
{\rm p.v.} \int\limits_{-b}^b\frac{\varphi(t)}{x-t}dt+\int\limits_{-b}^b{K(x,t)}\varphi(t)dt=f(x)
\end{equation}
can be rewritten as 
\begin{equation}
\label{eq:pvFx}
{\rm p.v.} \int\limits_{-b}^b\frac{\varphi(t)}{x-t}dt=F(x),
\end{equation}
where $F(x)=f(x)-\int\limits_{-b}^bK(x,t)\varphi(t)dt$. The solution of~\eqref{eq:pvFx}, in the case of a solutions limited at both extremes, is 
\begin{equation}\label{eq:phi_conclusion}
\varphi(x)=\frac{1}{\pi^2}\sqrt{b^2-x^2}\int\limits_{-b}^b\frac{F(t)}{\sqrt{b^2-t^2}}\frac{dt}{t-x}
\end{equation}
with the additional request 
\begin{equation}
\int\limits_{-b}^b\frac{F(t)}{\sqrt{b^2-t^2}}dt=0,
\end{equation}
which is  satisfied if $f(x)$ is an odd function, $\varphi(x)$ is an even function, and the kernel $K(x,y)$ transforms even functions into odd ones.
Applying this solution to~\eqref{eq:pvFx}, one has

\begin{align}
\nonumber
\varphi(x)&=\frac{\sqrt{b^2-x^2}}{\pi^2}{\rm p.v.}\int\limits_{-b}^b\frac{f(t)}{\sqrt{b^2-t^2}(t-x)}dt\\
&-\frac{\sqrt{b^2-x^2}}{\pi^2}{\rm p.v.}\int\limits_{-b}^b \frac{1}{\sqrt{b^2-t^2}(t-x)}dt\int\limits_{-b}^b K(t,\tau)\varphi(\tau)d\tau.
\end{align}
Inverting the order of integration in the second term, one has
\begin{align}
\nonumber
{\rm p.v.} \int\limits_{-b}^b\frac{1}{\sqrt{b^2-t^2}(t-x)}dt\int\limits_{-b}^b K(t,\tau)\varphi(\tau)d\tau \\
=
\int\limits_{-b}^b \varphi(\tau)d\tau  \left ( {\rm p.v.} \int\limits_{-b}^b \frac{K(\tau,t)}{\sqrt{b^2-t^2}(\tau-x)}dt \right)
\end{align}
and if one defines 
\begin{align}
&K_0(x,t)=\frac{1}{\pi^2}\sqrt{b^2-x^2} {\rm p.v.} \int\limits_{-b}^b \frac{K(\tau,t)}{\sqrt{b^2-\tau^2}(\tau-x)}d\tau\\
&f_0(x)=\frac{1}{\pi^2}\sqrt{b^2-x^2} {\rm p.v.} \int\limits_{-b}^b \frac{f(\tau)}{\sqrt{b^2-\tau^2}(\tau-x)}d\tau,
\end{align}
then~\eqref{eq:singular} has the same solution as the non-singular equation
\begin{equation}
\varphi(x)+\int\limits_{-b}^b K_0(x,t)\varphi(t)dt=f_0(t).
\end{equation}
The advantage of this transformation is that it transfers 
the calculation of the principal value of the equation to the calculation of the new known term and of the new kernel.

\bibliography{CS_polygonal_biblio}
\bibliographystyle{unsrt}

\end{document}